\documentclass[referee]{raa}            
\usepackage[driverfallback=dvipdfm]{hyperref}
\usepackage{graphicx,times}             
\usepackage{natbib}
\usepackage{amssymb,amsmath}
\bibpunct{(}{)}{;}{a}{}{,}

\hypersetup{colorlinks = true, linkcolor = green, anchorcolor = red, citecolor = blue, filecolor = red, pagecolor = red, urlcolor = red}

\begin{document}

   \title{First investigation of eclipsing binary KIC 9026766: analysis of light curve and periodic changes
}

   \volnopage{Vol.0 (20xx) No.0, 000--000}      
   \setcounter{page}{1}          

   \author{S. Soomandar
      \inst{1}
   \and A. Abedi
      \inst{1}
 
   }

   \institute{Faculty of Sciences, University of Birjand, Iran; {\it S.soomandar@birjand.ac.ir}\\
\vs\no
   {\small Received~~20xx month day; accepted~~20xx~~month day}}



%
%
\abstract{We investigate a short-period W UMa binary KIC 9026766 with an orbital period of 0.2721278 days in the Kepler field of view. By using an automated q-search for the folded light curve and producing a synthetic light curve for this object based on the PHOEBE code, we calculate the fundamental stellar parameters. We also analyze the O-C curve of the primary minima. The orbital period changes can be attributed to the combination of an upward quadratic function and light-travel time effect due to a possible third body with a minimum mass of $ 0.029M_{\bigodot_{}^{}}$ and an orbital period of $ 972.5866\pm0.0041 $ days.
The relative luminosity of the primary and secondary eclipses (Min I – Min II) is calculated. The periodogram of the residuals of the LTT effect, and Min I – Min II show peaks with the same period of  $ 0.8566 $ days. The background effect of two nearby stars on our target is the possible reason for this signal. By considering the amplitudes and periods of the remaining signals in the O-C curve of minima, spot motion is possible.
}

   \authorrunning{S. Soomandar \& A. Abedi }            
   \titlerunning{First investigation of eclipsing binary KIC 9026766: analysis of light curve and periodic changes}  
   \maketitle

%
\section{Introduction}           
\label{sect:intro}
The role of eclipsing binary stars (EBs) is very important in stellar astrophysics. Properties such as the mass, radius, and luminosity of these stars can be calculated with an accuracy of less than $1$ percent (\citealt{Pietrzynski+etal+2019}), and the existing stellar evolution models can be modified using these parameters (\citealt{Torres+etal+2014}).
Eclipsing binary stars (EBs) contain two stars where the components orbit around their common barycenter (i.e., their orbit results in eclipses), and most of our knowledge about their characteristics comes from this motion. One of the reliable measures to determine the parameters of the components is the variations in the O-C curve of minima. Physical factors such as mass transfer (\citealt{Pribulla+etal+2000}), apsidal motion (\citealt{Guinan+Maloney+1985}), Applegate effect (\citealt{Applegate+1992}), Shklovskii effect (\citealt{Shklovskii+1970}), Barycentric and Asymmetric Transverse Velocities (BATV) (\citealt{conroy+etal+2018}), light-travel time effect (\citealt{Borkovits+Hegedus+1996}), and dynamical effects of a third body on the binary orbit can be the reasons for the variations in the O-C curve.

In addition to these effects, the observed erratic variations may indicate unidentified apparent timing effects (\citealt{Borkovits+etal+2015}). Furthermore, stellar spots can cause the spurious O-C curve variations, which are completely investigated in \cite{Tran+etal+2013, Balaji+etal+2015}.

KIC 9026766 is a contact binary in the Kepler field of view, and one of its flags in the of eclipsing binary catalog (\cite{abdul2016})\footnote{http:// keplerebs.villanova.edu/} is TM that indicates multiple bodies in the system. Therefore, we analyze the O-C curve to investigate the triple-body effect of the binary. Moreover, the light curve is modeled.

\section{Observations}
\label{sect:Obs}

KIC 9026766 was observed by Kepler satellite during quarters 0-17 in long-cadence (LC) mode with effective integration times of 30 minutes. No light curve has been reported for this target in short-cadence (SC) mode (\citealt{abdul2016}). 
Some parameters are presented in the Kepler EB catalog, such as the morphology parameter which was introduced by \cite {Matijevic+etal+2012}
as an identifier to determine the type of binary that has the value between 0 and 1, where 0 indicates a completley detached binary, and 1 indicates a contact binary. Moreover, the effective temperature (accurate to 200 K) of the target is estimated using ground-based optical multi-color photometry of the Two Micron All Sky Survey (2MASS) (\citealt{Skrutskie+etal+2006}). Other observational parameters such as surface gravity Log g (accurate to 0.5 dex), Barycentric Julian date of the primary eclipse BJD0 (accurate to 0.015238), and the period of the binary (accurate to 0.0000002 days) are presented in Tabel ~\ref{Tab1}. The nearby stars Gaia DR2 2127972818460821504 and Gaia DR2 212 7972818467257344 are located within 0.5 and 1.0 arcsec of the target\footnote{http://cdsportal.u-strasbg.fr/?target=KIC 209026766}. Since each pixel is 4 arcsec across, the blending of sources is expected. We checked the pixel data using Lightkurve Python package (\citealt{Collaboration+2018}) for each quarter of the chosen target pixel files in the aperture; \cite{abdul2016} was selected the best pixel data for this system, so we used the detrended light curves in the Kepler EB catalog.
\begin{table}
\begin{center}
\caption[]{ Observation Specifications of KIC 9026766.}\label{Tab1}
 \begin{tabular}{clcl}
  \hline\noalign{\smallskip}
Parameter &  Value in Catalog    \\
  \hline\noalign{\smallskip}
Kepler ID&  9026766    \\ 
2Mass ID  &J19341881 + 4518354  \\                 
RA  & 19 34 18.816   \\
Dec & + 45 18 35.46  \\
Kepler mag. & $ 14.998\pm 0.02$\\
$T_{eff}(K)$ & $4888\pm2.0$\\
$log(g)$& $4.7\pm0.5$\\
Period(days) & $ 0.2721278\pm2E-7$\\
$Bjd_{0}$& $2454964.846798\pm0.015238$\\
$Morphology$ & $ 0.76$\\
  \noalign{\smallskip}\hline
\end{tabular}
\end{center}
\end{table}
\section{Light Curve Solution}
\label{sect:data}

To find the initial guess of the orbital parameters of KIC 9026766, we performed an automated q-search in the same way done by \cite{Soomandar+Abedi+2020}. The Wilson-Devinney code was imported as a subroutine to Matlab, and the least square method was used to obtain the initial guess of the parameters.

\begin{figure}
   \centering
   \includegraphics[width=\textwidth, angle=0]{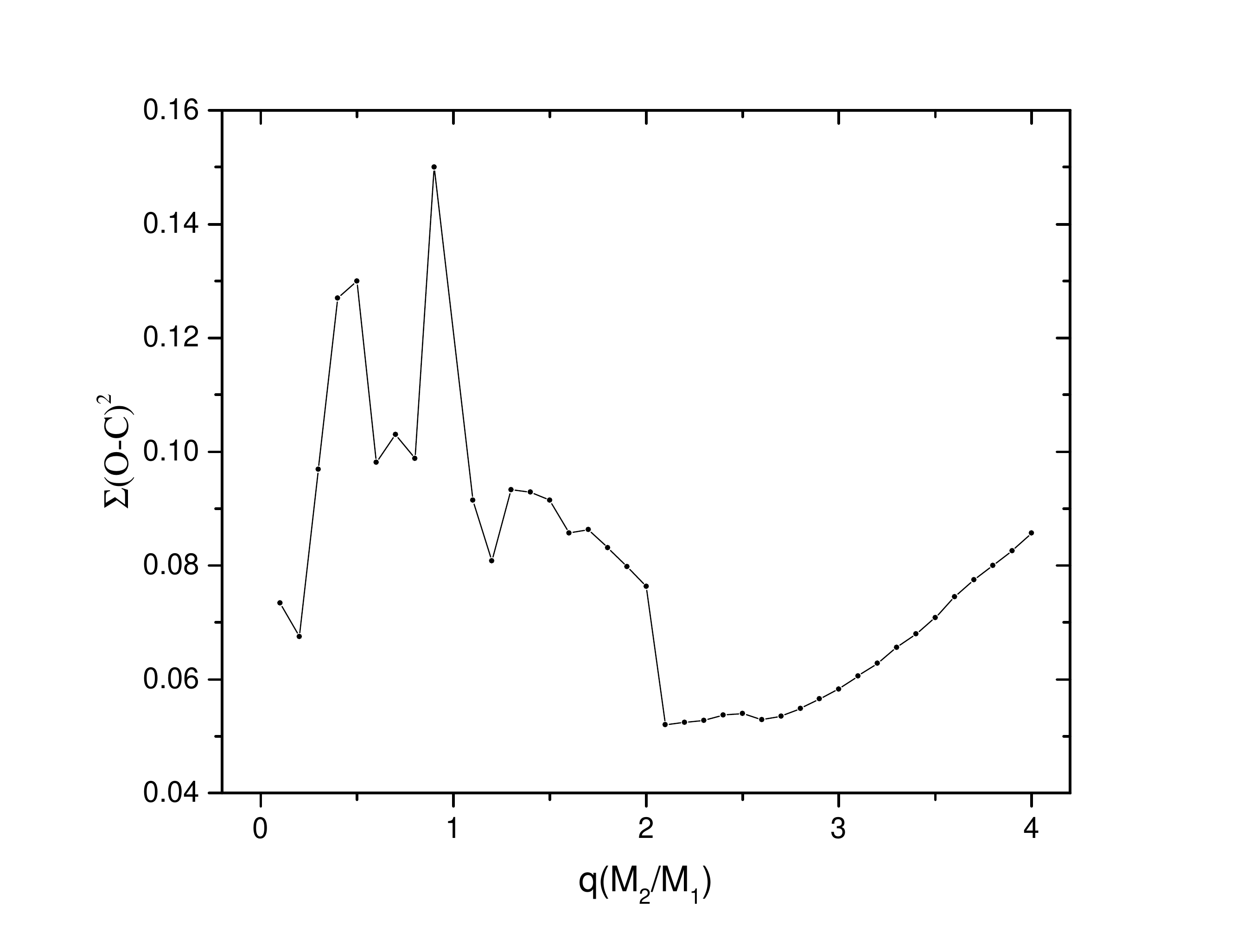}
   \caption{Automated q-search in the range [0.1-4.00].}
   \label{Fig1}
   \end{figure}

One cycle of the light curve in LC mode has almost 14 data points. For the light curve solution, we used 4.08 days data points and phase-binned them.
The morphology parameter of this target was reported as 0.78, so we chose mode 3 (contact not in thermal equilibrium) in the Wilson-Devinney code. The fixed parameters are as follows: effective temperature, $T_{1}=T_{eff}$, bolometric albedos $ A_{1}=A_{2}=0.5 $ (\citealt{Rucinski+1969}), and gravity brightening $g_{1}=g_{2}=0.32$ (\citealt{Lucy+1967}). Moreover, we chose the linear regime of \cite{van+1993} for limb darkening coefficients and considered two reflection effects. For q, the iteration was considered within the range [0.1-4.00] with 0.1 steps; for the temperature of the secondary component, we iterated within [4000-7000] with $1000^{0}K$ steps, and the stellar gravitational potential parameter $\Omega_{1}=\Omega_{2}$ was iterated in the range $[\Omega(L_{1}) -\Omega(L_{2})]$ with 0.05 steps. The minima of indicate that $\Sigma(O-C)^{2}$for q=2.1, the corresponding curve is plotted in Figure \ref{Fig1}. \cite{Prsa+etal+2011} \footnote{http://keplerebs.villanova.edu/v2} catalog lists the principal parameters determined by a neural network analysis of the phased light curves. One of the computed parameters of contact systems is the photometric mass ratio, and our result for q=2.1 is close to what they have calculated q=2.09477. Therefore, we chose this value for the light curve solution.

\begin{figure}
   \centering
   \includegraphics[width=\textwidth, angle=0]{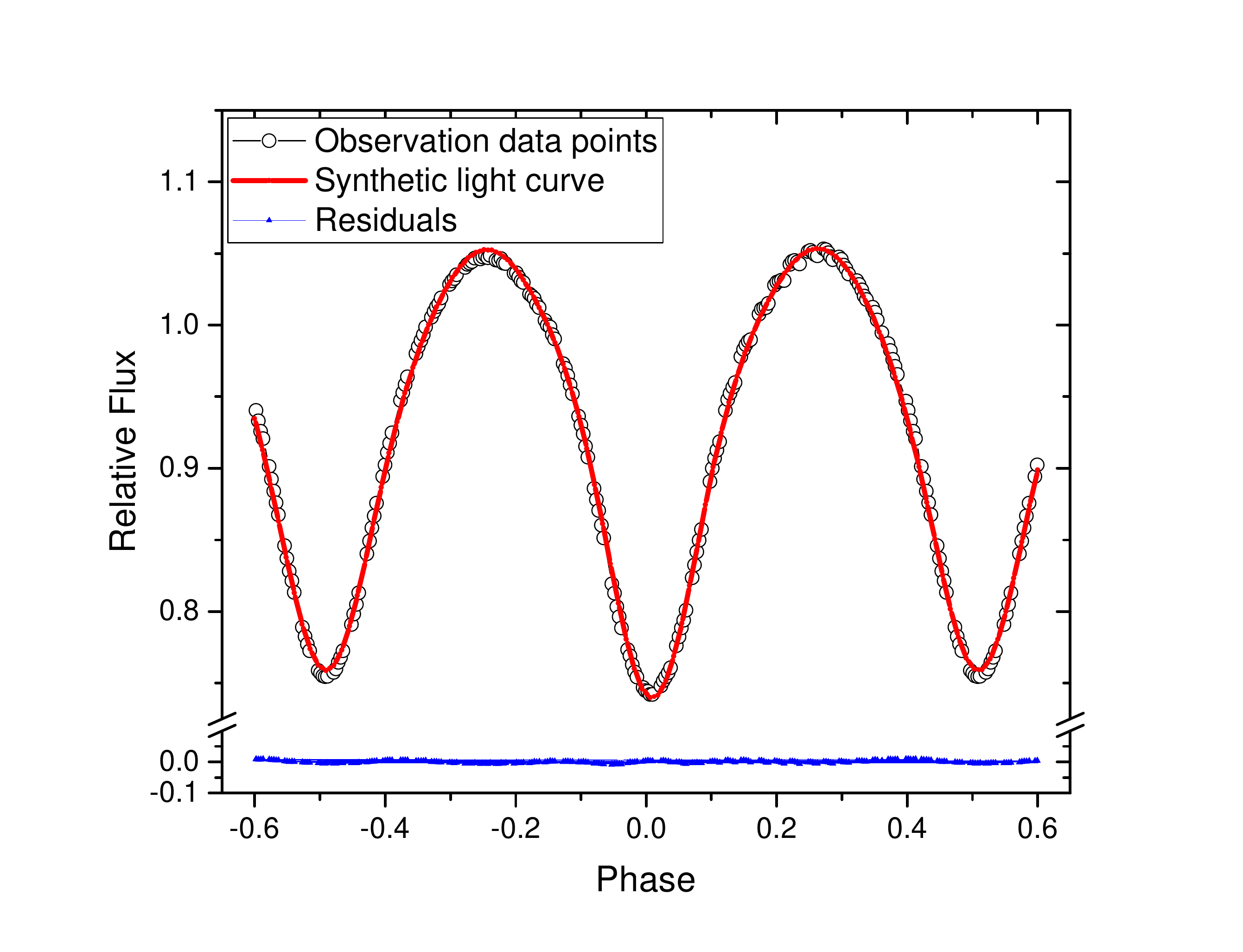}
   \caption{ Light curve solution of KIC 9026766. Observational data points (blank circles), synthetic light curve (solid red line), residuals (blue triangles).}
   \label{Fig2}
   \end{figure}

\begin{figure}[h]
  \begin{minipage}[t]{0.495\linewidth}
  \centering
  
   \includegraphics[width=\textwidth, angle=0]{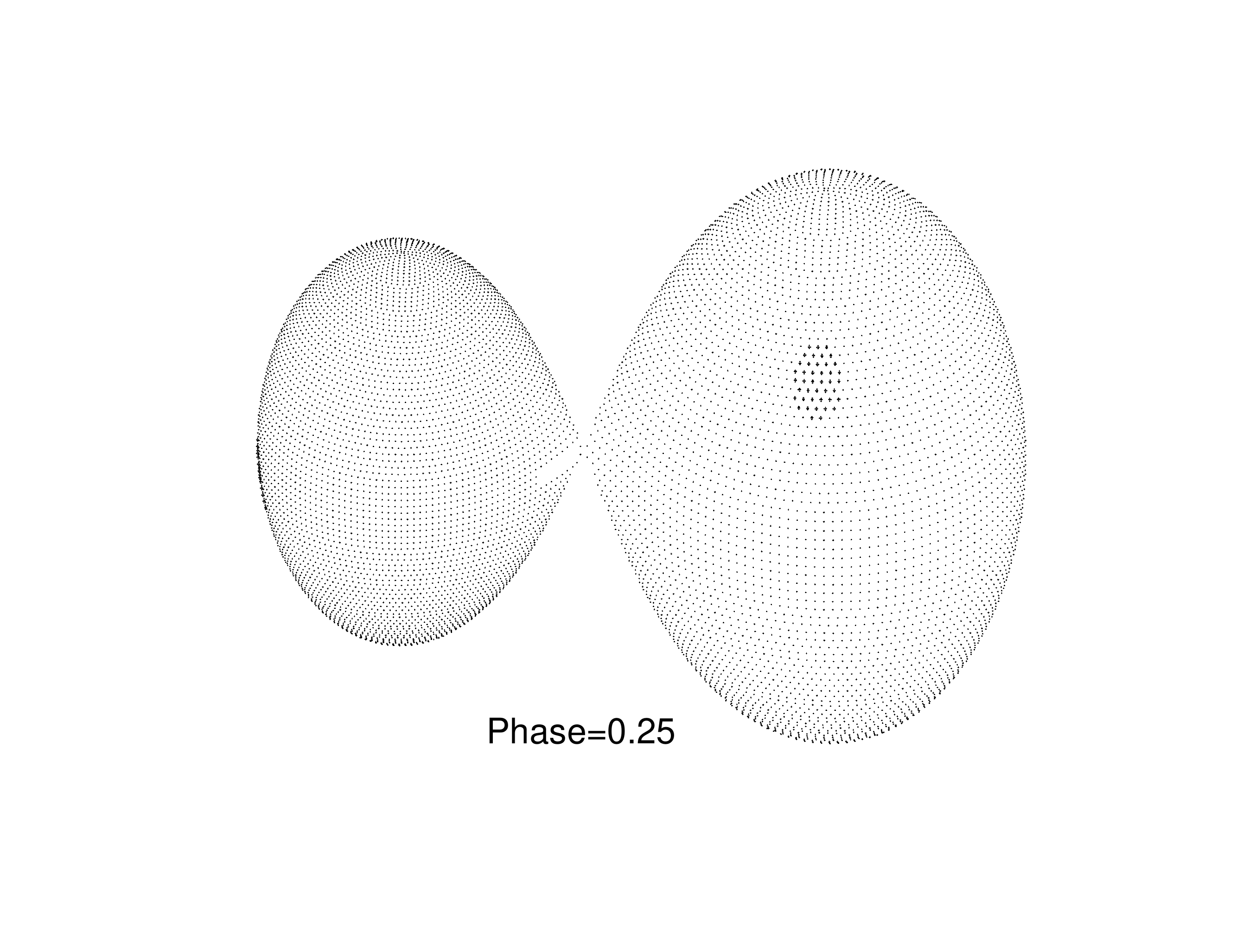}
   \centering
   
  \end{minipage}%
  \begin{minipage}[t]{0.495\textwidth}
  \centering
   \includegraphics[width=75mm]{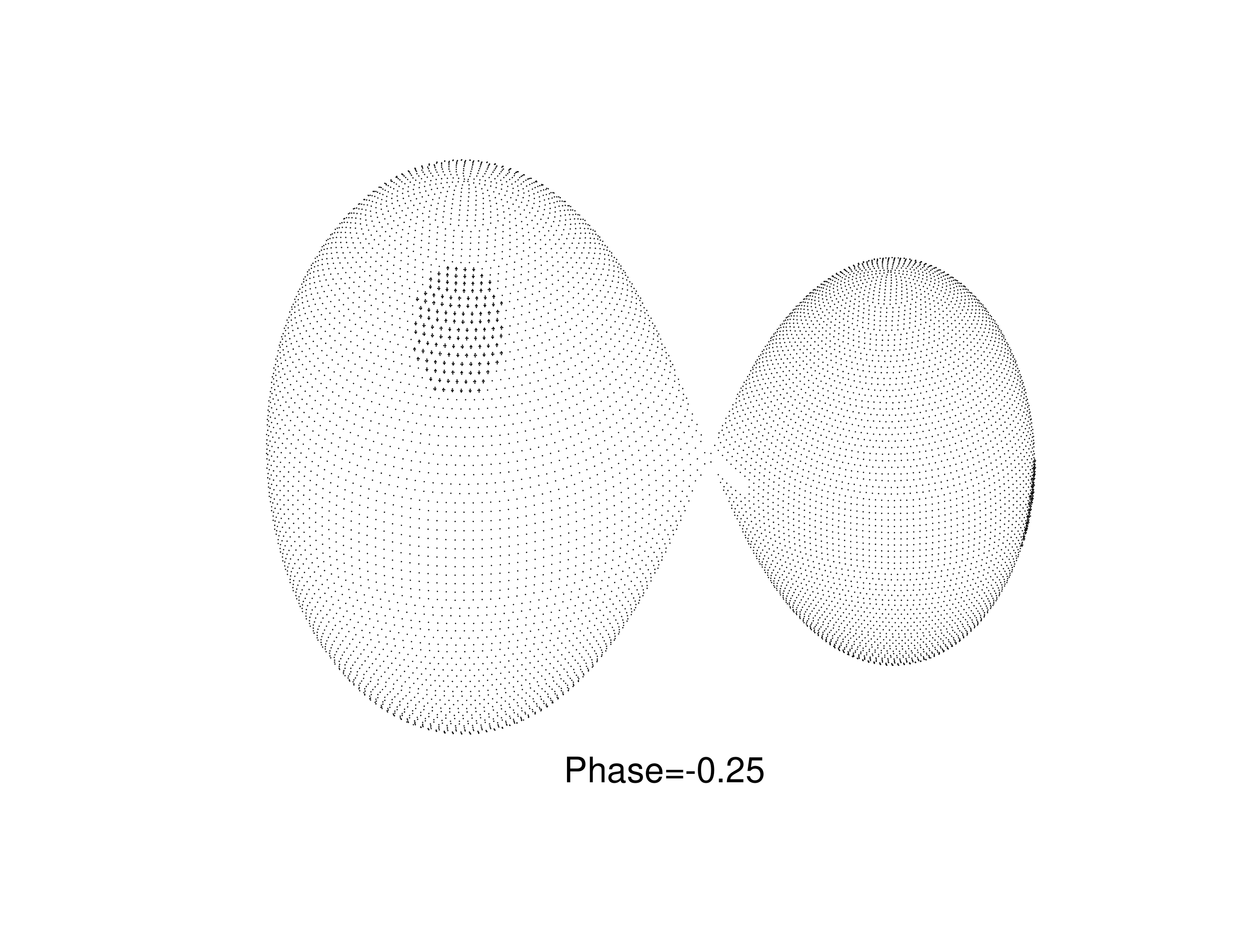}
  \end{minipage}%
  \caption{Three-dimensional view of KIC 9026766 at phases -0.25,0.25}
  \label{Fig3}
\end{figure}

In the following, we use the PHOEBE Legacy (SVN 2012-07-08) by selecting mean Kepler passband and using finite integration time for LC mode.
We repeated fitting by considering the differential correction method, and we used the following five free parameters: the temperature of the second star,$ T_{2}$, orbital inclination, i, stellar surface gravitational potential, $\Omega_{1}=\Omega_{2}$ , the monochromatic luminosity of the primary star, $ L_{1} $, and q. 
Then, we added cool spots characterized by radius, temperature, co-latitude, and latitude on the surface of the two components to obtain the best solution. The final parameters are tabulated in Table \ref{Tab2}. Figure \ref{Fig2} shows the synthetic light curve, and Figure \ref{Fig3} shows the three-dimensional model of the system.


\begin{table}
\begin{center}
\caption[]{ Orbital Parameters of Light Curve Solution for KIC 9026766.}\label{Tab2}

 \begin{tabular}{clcl}
  \hline\noalign{\smallskip}
Parameter &  Value   \\
  \hline\noalign{\smallskip}
$ A_{1}=A_{2}^{a}$ &  0.5    \\ 
$g_{1}=g_{2}^{a}$ &0.32  \\                 
$T_{1}^{a}$ & 4880   \\
$T_{2}$& $4757\pm0.411$ \\
q & $ 2.077\pm 0.0068$\\
$i^{0}$ & $69.42\pm0.04$\\
$\Omega_{1,2}$& $5.3591\pm0.0006$\\
$ \frac{L_{1}}{L_{1}+L_{2}}$ & $ 0.3692\pm0.0041$\\
$f$ & $0.2$ percent\\
$x_{1}^{a}$ & $ 0.66025$\\
$x_{2}^{a}$ & $0.66699$\\
$r_{1}(pole)$ & $ 0.2961\pm0.084$ \\
$r_{2}(pole)$& $0.4174\pm0.008$\\
$r_{1}(side)$& $0.3098\pm0.099$\\
$r_{2}(side)$& $0.4447\pm0.105$\\
$r_{1}(back)$& $0.34243\pm0.153$\\
$r_{2}(back)$& $0.47142\pm0.136$\\
Spot $co-latitude^{b}$ & $100.01\pm0.015$ \\
Spot $longitude^{b}$& $181.20\pm0.02$ \\
Spot $radius^{b}$& $ 12.01\pm0.02$\\
$(\frac{T_{spot}}{T_{surf}})^{b}$ & $0.9\pm0.02$\\
Spot $co-latitude^{c}$& $45.01\pm0.025$\\
Spot $longitude^{b}$&$271.20\pm0.02$\\
Spot $radius^{b}$& $ 15.2\pm0.1$\\
$(\frac{T_{spot}}{T_{surf}})^{c}$& $0.95\pm0.02$\\
Spot $co-latitude^{c}$& $55.01\pm0.023$\\
Spot $longitude^{b}$&$90.0\pm0.03$\\
Spot $radius^{b}$& $ 8.1\pm0.1$\\
$(\frac{T_{spot}}{T_{surf}})^{c}$& $0.8\pm0.02$\\
$\Sigma(O-C)^{2}$& 0.0321\\
  \noalign{\smallskip}\hline
\end{tabular}
\end{center}
\tablecomments{0.86\textwidth}{$^{a}$: Fixed Parameter $^{b}$: Primary Star $^{c}$: Secondary Star.}
\end{table}
To calculate the mass of the components of KIC 9026766, we assume that more massive component is a normal main sequence star, its mass could be estimated as $M_{2}=0.73M_{\bigodot_{}^{}}$, corresponding to the spectral type K2 (\citealt{Cox+2000}). By using the mass ratio q and $M_{2}$ values, the mass of primary component is calculated as $M_{1}=0.35 M_{\bigodot_{}^{}}$.  

\section{Measuring Minima}

The observed minima were calculated by fitting a Lorentzian function to the individual minima in the same way done by \cite{Soomandar+Abedi+2020}, To find the initial parameters of Lorentzian function we used the maximums and minimum relative flux and mean Barycentric Juilan date for individual eclipse. And we used $Scipy.curve-fit$  package of Python to minimize $\Sigma(O-C)^{2}$ and calculated the observational time eclipses. 

The linear ephemeris of this target is reported on keplerebs.villanova.edu website as follows:
\begin{equation}
T_{min,1}=2454964.8467981(\pm0.015238)+0.2721278(\pm0.0000002)E                       
\end{equation}
             
\begin{figure}
   \centering
   \includegraphics[width=\textwidth, angle=0]{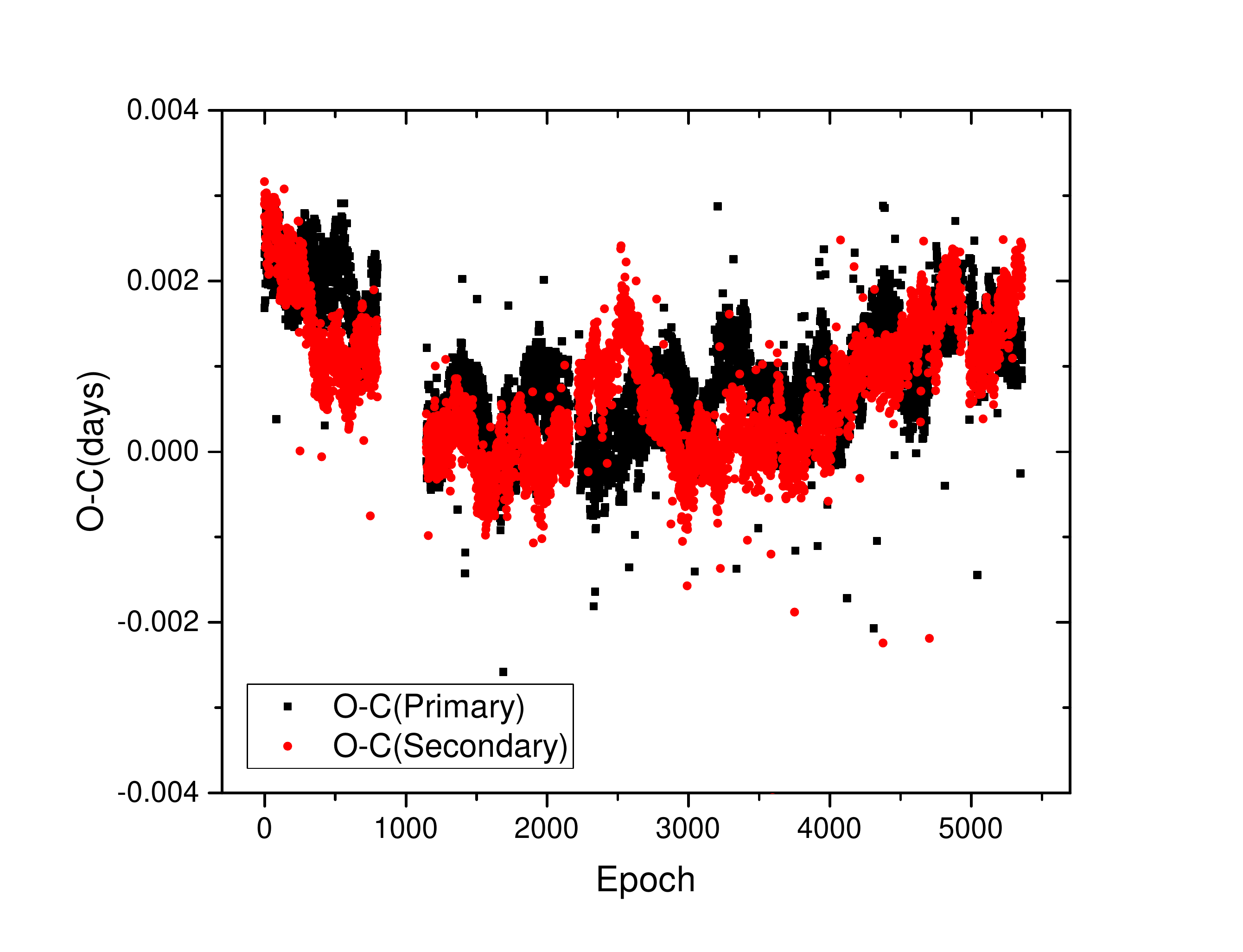}
   \caption{ O-C curves of minima of KIC 9026766. The primary and secondary minima of eclipses are represented by black squares and red circles, respectively.}
   \label{Fig4}
   \end{figure}

Where E is the observational epoch. We used Eq. 1 to calculate the O-C curves of the primary and secondary minima of eclipses.
The O-C curves of minima are plotted in Figure ~\ref{Fig4}.

\begin{figure}
   \centering
   \includegraphics[width=\textwidth, angle=0]{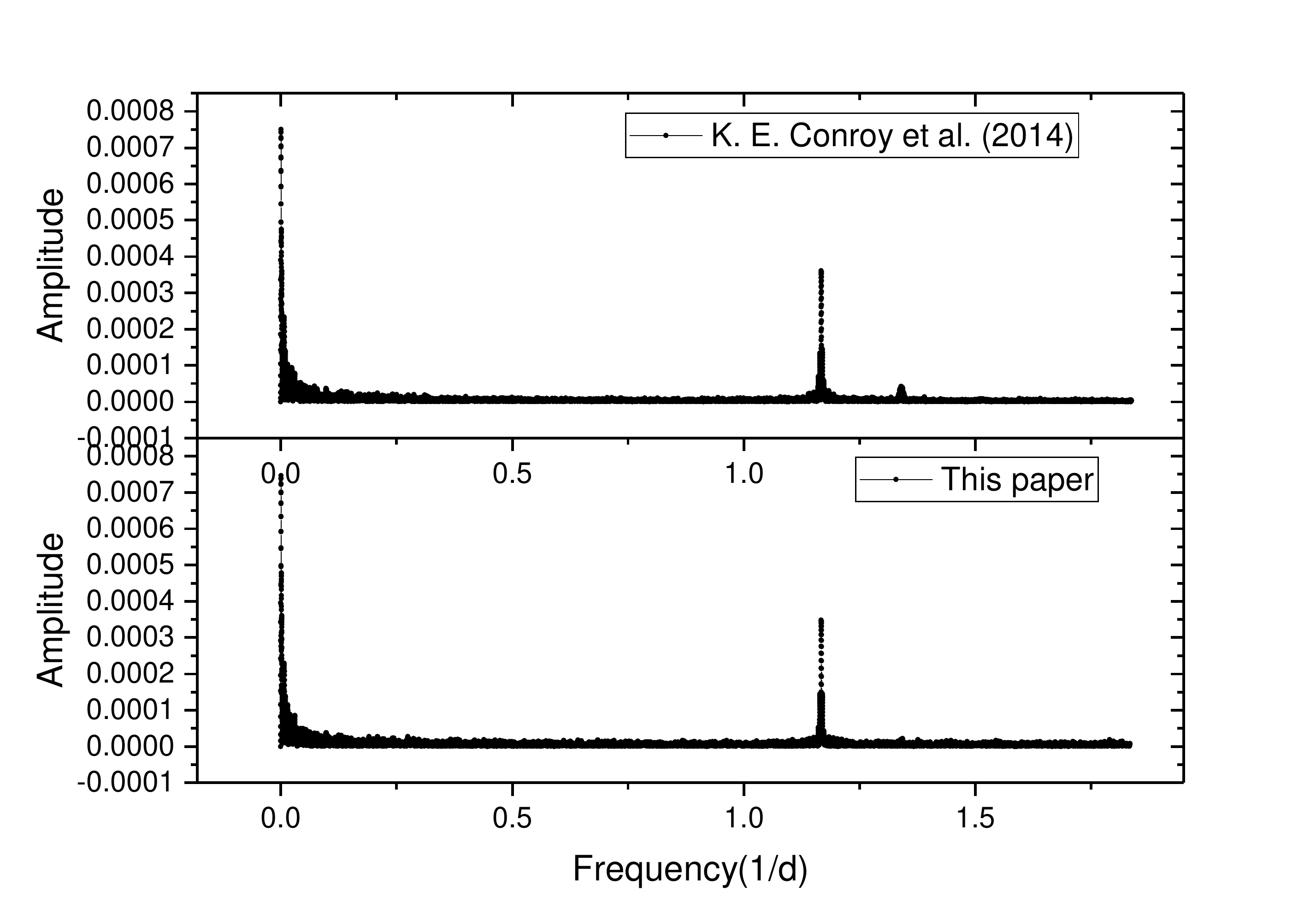}
   \caption{Top panel: The periodogram analysis of the O-C curve of \citealt{conroy+etal+2014}. Bottom panel: The periodogram analysis of O-C curve of this work.}
   \label{Fig5}
   \end{figure}

\cite{conroy+etal+2014} calculated the precise eclipse times for 1279 close binaries in the Kepler eclipsing binary catalog. To compare the result of this work to the calculated O-C curve by \cite{conroy+etal+2014}. We downloaded the O-C curve of primary minima for this target and compared the periodogram of their results with the periodogram of primary minima of this work.  The result curve is shown in the Figure ~\ref{Fig5}. 
\section{Orbital Period Changes}
\begin{figure}[h]
  \begin{minipage}[t]{0.5\linewidth}
  \centering
   \includegraphics[width=75mm]{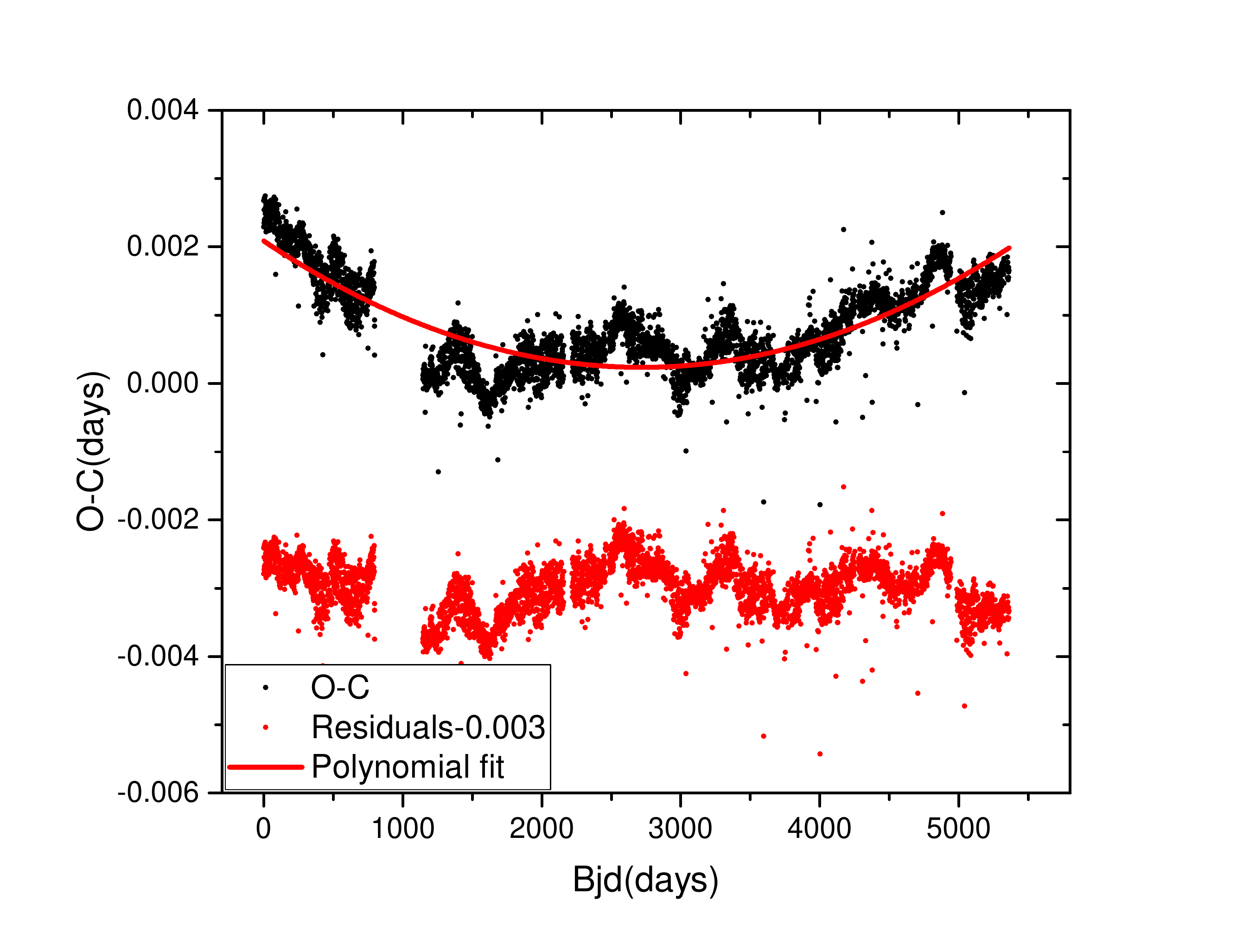}
  \end{minipage}%
  \begin{minipage}[t]{0.5\textwidth}
  \centering
   \includegraphics[width=75mm]{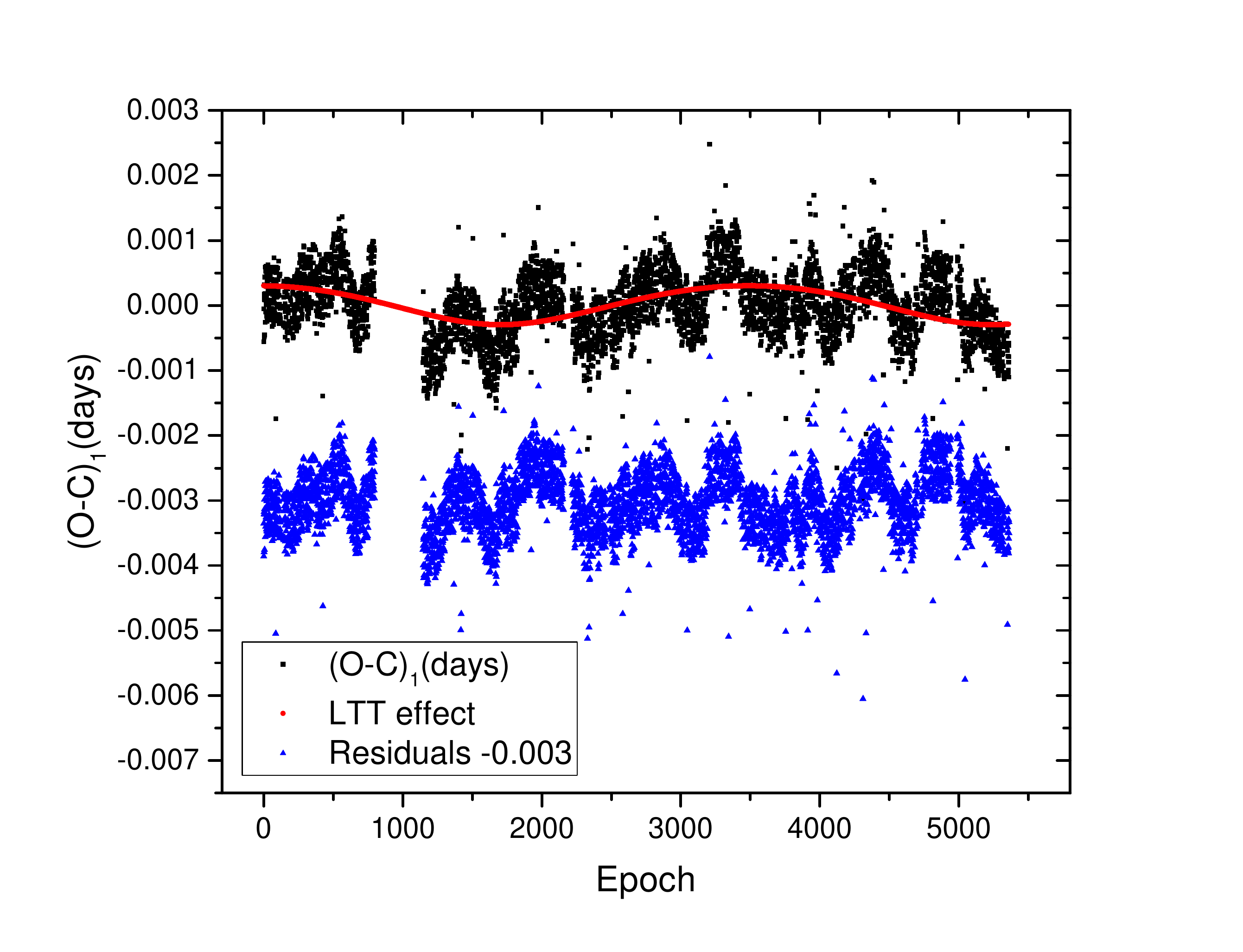}
  \end{minipage}%
  \caption{Left panel: fitting the quadratic function to the O-C curve of the primary minima. O-C curve of the primary minima (black squares), quadratic function fit (solid red line), residuals of this fitting (red circles). Right panel: LTTE on the residuals of quadratic fit. Residuals of quadratic fit (black squares), LTTE (red circles) and residuals of LTTE (Blue triangle).}
  \label{Fig6}
\end{figure}

To investigate the orbital changes over time, we considered the O-C curve of the primary minima. Fitting a quadratic function to the O-C curve of minima exhibits an increase in the period; the corresponding plot is shown in the left panel of Figure  ~\ref{Fig6}, and the fitting coefficients are listed in Table~\ref{Tab3}.

\begin{table}
\begin{center}
\caption[]{ Coefficients of Quadratic Fit.}\label{Tab3}

 \begin{tabular}{clcl}
  \hline\noalign{\smallskip}
A &  B   & C \\
  \hline\noalign{\smallskip}
$0.00224\pm2.35724\times10^{-5}$&  $-1.35557\times10^-6\pm1.96721\times10^{-8}$ &$2.42862\times10^{-10}\pm3.54376\times10^{-12}$   \\ 

  \noalign{\smallskip}\hline
\end{tabular}
\end{center}
\end{table}
The rate of period increase is calculated by Eq. 2 (\citealt{Hilditch+2001}).
\begin{equation}
\dot{P}=\frac{2C}{P_{orb}}=\frac{2\times2.42862\times10^{-10}}{0.2721278}=1.784911\times10^{-9} \pm7.08752\times10^{-12}\frac{day}{year}
\end{equation}
By considering mass conservation, the mass exchange rate can be calculated using Eq. 3. 
\begin{equation}
\frac{\dot{P}}{P}=-3\frac{\dot{M_{2}}(M_{1}-M_{2})}{M_{1}M_{2}},\dot{M_{2}}=+1.4700\times10^{-9}\pm0.5837\times10^{-11}\frac{M_{\bigodot_{}^{}}}{year}
\end{equation}

where the positive sign indicates that the direction of mass transfer is from the less massive component to the more massive one. 

The residuals of quadratic fit show periodic changes over time, so we consider the light-travel time effect (LTTE) as the possibility of the presence of a third body.

In the following, we apply the periodogram analysis by using Period04
(\citealt{Lenz+Breger+2005}) for the residuals of the quadratic fit. The strong peak shows a period of almost 972 days for the third body. The LTTE term is given in Eq. 4 (\citealt{Irwin+1952}).

\begin{equation}
(O-C)_{LTTE}=\frac{A}{\sqrt{1-e^{2}\cos^{2}(\omega)}}(\frac{1-e^{2}}{1+e\cos(\nu)}\sin(\nu+\omega)+e\sin(\omega))
\end{equation}

where $A=\frac{a_{1,2}\sin(i)}{c}\sqrt{1-e^{2}\cos^{2}(\omega)}$ and $a_{1,2}$is the semi-major axis of the relative orbit of the eclipsing system around the center of mass (in au), $i$ is the inclination of the third-body orbit, $e$ is the eccentricity of the third-body orbit, $ \omega$ is the longitude of periastron, and $\nu $is the true anomaly. The calculated parameters of the third body enable us to determine the mass function $ f_{m}(M_{\bigodot_{}^{}})$ of the triple system. By assuming the coplanar orbit ($i=90^{0}$), we obtained the lower limit for the mass of the third component. The derived parameters of LTT solution are listed in Table ~\ref{Tab4} and the resulting curve for the third-body effect is plotted in the right panel of Figure~\ref{Fig6}. 
\begin{table}
\begin{center}
\caption[]{ Parameters of the Third Body Around KIC 9026766.}\label{Tab4}

 \begin{tabular}{clcl}
  \hline\noalign{\smallskip}
Parameter &  Value For the Third Body   \\
  \hline\noalign{\smallskip}
Eccentricity (e)&  $0.2\pm0.005$    \\ 
$Omega(\omega)$  &$259.2\pm1.8$  \\
                 
A (the semi-amplitude)(days) & $3.0E-4\pm1.5E-4$  \\
$T_{0}$ & 2456372.0000 \\
Projected semi-major axis,$ a_{1,2}\sin(i) (au)$ & $ 5.197E-2\pm 2.598E-2$\\
Mass Function,$ f_{m}(M_{\bigodot_{}^{}})$& $1.98E-5\pm0.9E-9$\\
Period of the third body(days) & $ 972.5866\pm0.0041$\\
$M_{3}sin(i)(M_{\bigodot_{}^{}},i=90)$& $0.029$\\

  \noalign{\smallskip}\hline
\end{tabular}
\end{center}
\end{table}
By considering the estimated mass of the third body, a brown dwarf is possible. 

\begin{figure}
   \centering
   \includegraphics[width=\textwidth, angle=0]{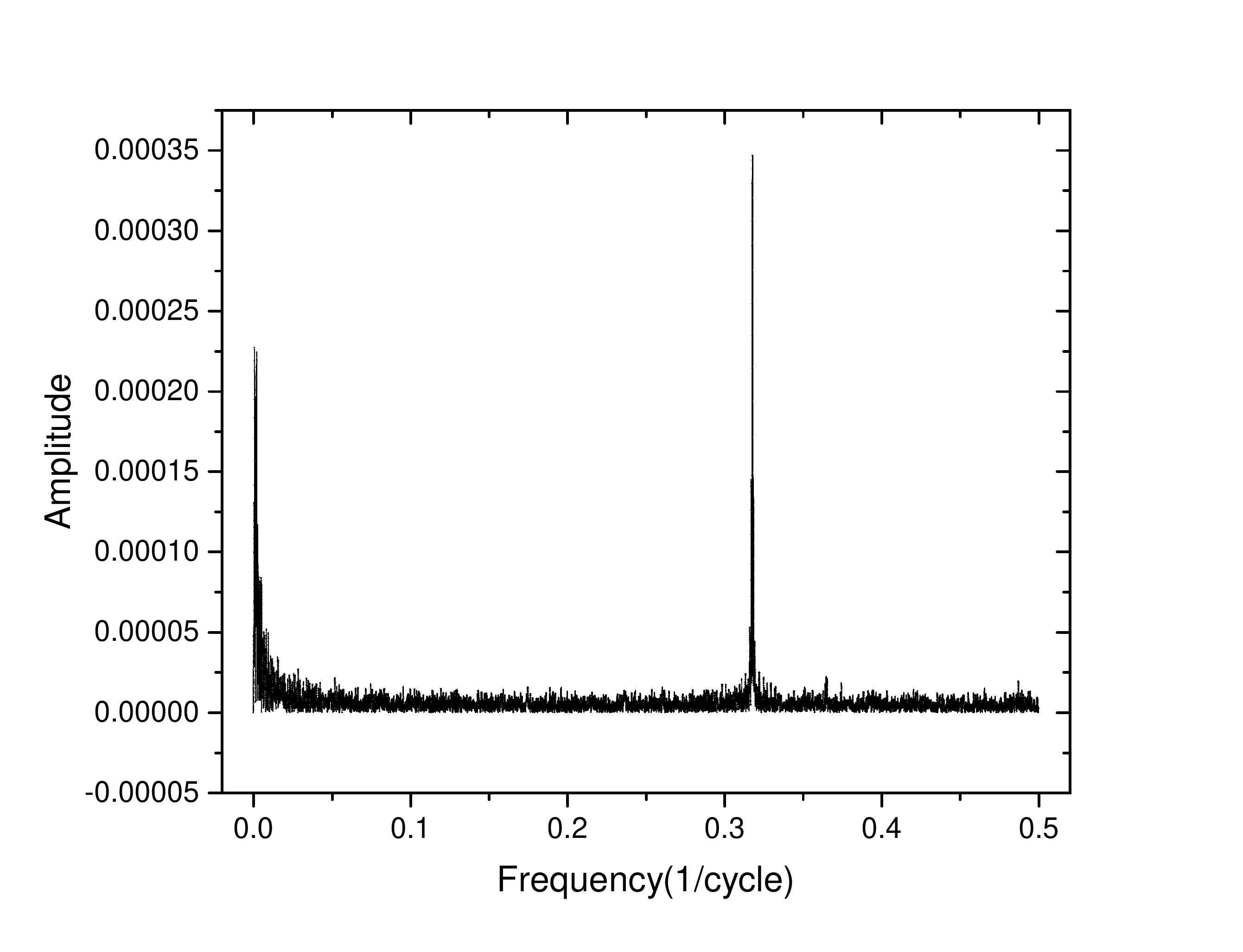}
   \caption{ Periodogram of residuals after removing the LTT effect.}
   \label{Fig7}
   \end{figure}

\begin{figure}[h]
  \begin{minipage}[t]{0.5\linewidth}
  \centering
   \includegraphics[width=75mm]{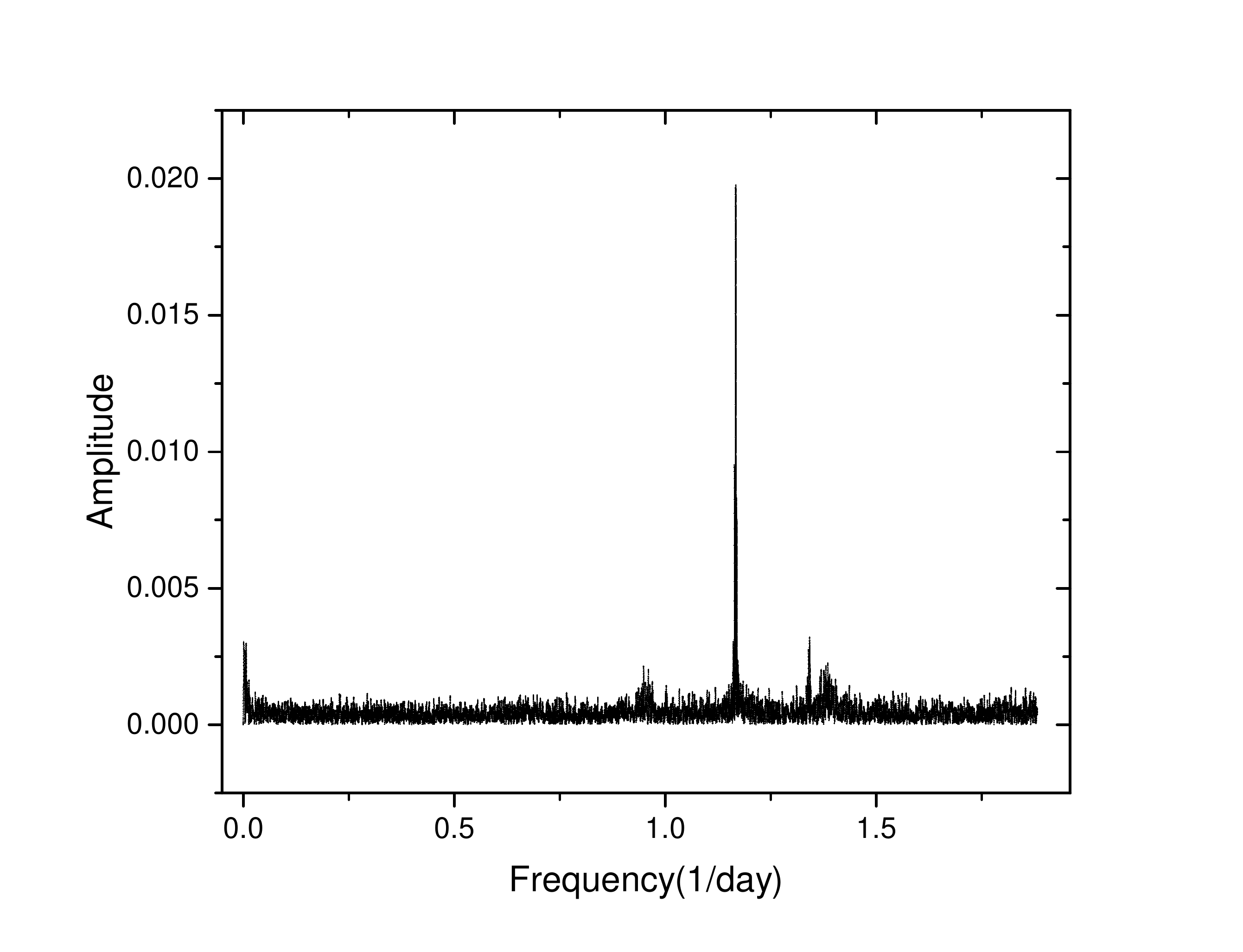}
  \end{minipage}%
  \begin{minipage}[t]{0.5\textwidth}
  \centering
   \includegraphics[width=75mm]{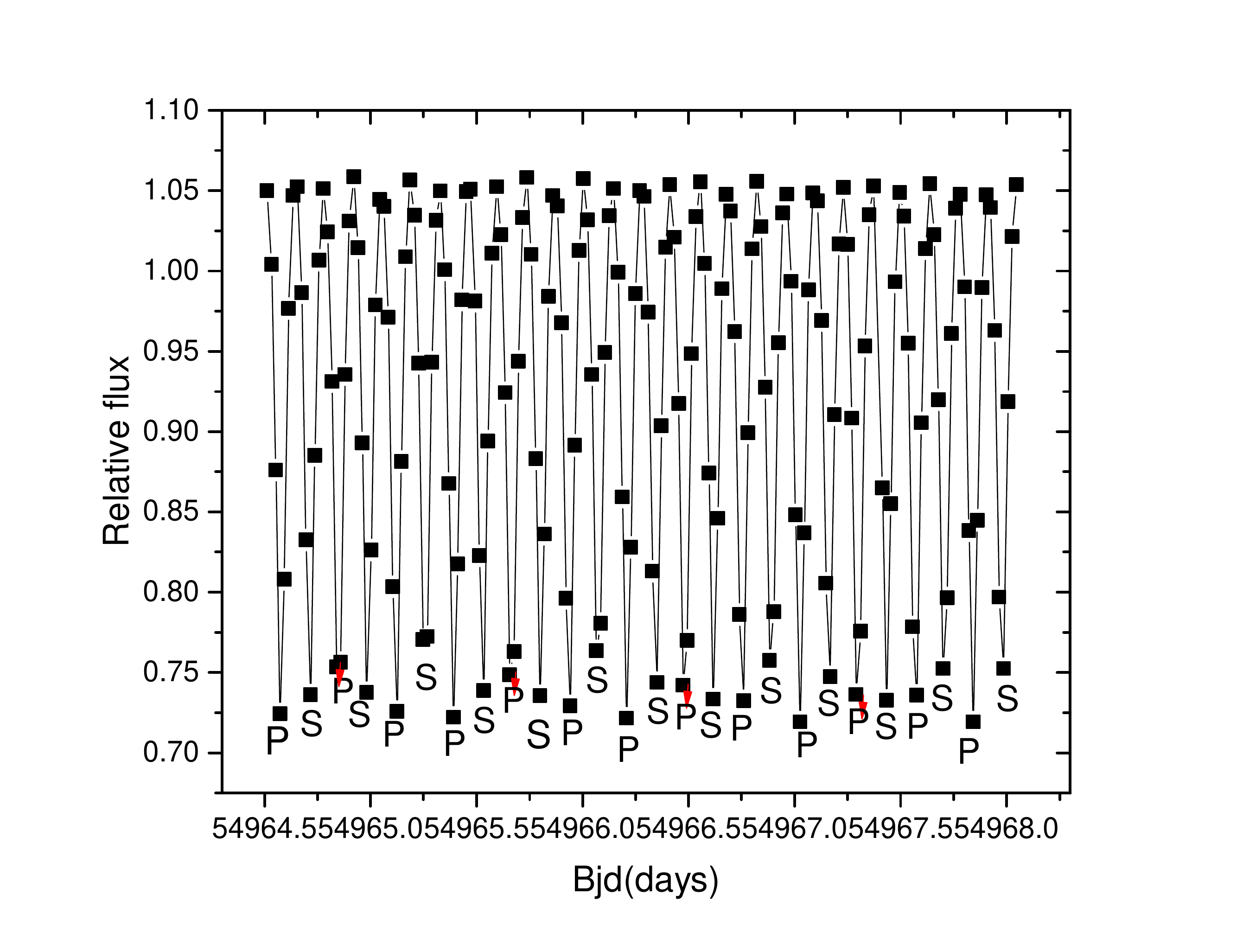}
  \end{minipage}%
  \caption{Left panel: periodogram of the relative luminosity of the primary and secondary eclipses (Min I – Min II). Right panel: light curve over two days; primary minima are denoted by P and secondary ones by S; red arrows show when the depth of the primary minima is less than that of the secondary ones.}
  \label{Fig8}
\end{figure}

In the following, we apply Period04 (\citealt{Lenz+Breger+2005}) to the residuals of LTT effect. The periodogram is plotted in Figure~\ref{Fig7}. The highest peak in the periodogram has a period of $0.8566$ days. 
Thus, we calculated the relative luminosity of the primary and secondary eclipses (Min I – Min II) for every light curve over 4 years. The peak frequency is $f=1.16738\pm9.8133e-6(\frac{1}{d})$, which is equal to a period of 0.8566 days. In the right panel of Figure~\ref{Fig8}, the light curve for 2 days is plotted and red arrows show when the depth of the primary minima is less than that of the secondary ones. This occurs almost every 0.8566 days, so the reason for this short periodicity is the exchange between the primary and secondary minima, which affects the O-C curve of minima.
Another possible reason for the cyclic variations in the O-C curve of minima is the Applegate effect. The magnetic field of stars could produce angular momentum transfer among its internal and external layers, causing orbital period changes \citealt{Applegate+1992}. 
\begin{equation}
\frac{\Delta(P)}{P}=2\pi\frac{(O-C)}{P_{mod}}
\end{equation}
where (O-C) is the amplitude of the O-C curve and $P_{mod}$ is the luminosity period changes. Having the same period as the O-C period modulation is necessary. To calculate the luminosity period changes, we considered the flux in the first and second maxima of individual light curves over 4 years. Then, we calculated the period modulation for the difference in the primary and secondary maxima; $P_{mod}=0.8566$ days, so $ \frac{\Delta(P)}{P}\approx2.538\times10^{-3}$. This short periodic oscillation and the value of $\frac{\Delta(P)}{P}$suggests that the Applegate mechanism cannot explain the cyclic variations in the O-C curve.

As mentioned in Section 2, the two nearby stars located within 0.5 and 1 arcsec of our target can cause background effects, which may not be noticed in the pixel data. This can cause a signal with a period of just more than three times the orbital period. 
The spot evolution in contact binaries can affect the eclipse times and mimic changes in the orbital period (\citealt{Tran+etal+2013}). In the following, we consider the spot evolution as the main reason for changes in the O-C curves of minima.

To examine the effect of spots on the O-C curve of minima, we used the theoretical formula derived by \cite{Pribulla+etal+2012} to estimate the amplitude ($\sim 0.0004days$) of eclipse timing variations, which are sorted by the remaining frequency peaks (remained after removing the third-body peak) in the periodogram of the O-C curve (see ~Table\ref{Tab5}). Additionally, the observed quasi-periods of 132-200 days are in good agreement with periods of 50-200 days that were observed for contact binaries in Kepler field of view (\citealt{Tran+etal+2013}). The remaining frequency peaks in the periodogram of the O-C curve are listed in Table ~\ref{Tab5}.
\begin{table}
\begin{center}
\caption[]{ Remaining Frequency Peaks in O-C Curve.}\label{Tab5}

 \begin{tabular}{clcl}
  \hline\noalign{\smallskip}
Frequency(1/cycle) &  Amplitude(days) & Period(days)    \\
  \hline\noalign{\smallskip}
$f_{1}=0.0020518\pm3.06486e-6$& $ 0.0002258\pm6.610e-6$ & $132.6288\pm0.1982 $     \\ 
$f_{2}=0.0007368\pm3.06486e-6$  &$ 0.0002216\pm6.610e-6$   &  $ 369.3374\pm1.5333$\\                 
$ f_{3}=0.001679\pm3.37546e-6$  &$ 0.0001782\pm6.610e-6$    &   $ 169.0733\pm0.3257$ \\
$f_{4}=0.001380\pm3.7932e-6$ & $ 0.0001626\pm 6.610e-6$ & $ 197.1940\pm0.5410$   \\
  \noalign{\smallskip}\hline
\end{tabular}
\end{center}
\end{table}
\section{Discussion and Conclusion}
\label{sect:Discussion and Conclusion}
We studied a short-period W UMa binary, KIC 9026766, in the Kepler field of view. Our study had two parts. In the first part, we derived the fundamental stellar parameters of the light curve using PHOEBE Legacy. For this purpose, we performed an automated q-search and found the mass ratio of the binary ($q=2.077\pm0.0068$). We assumed the more massive is a main sequence, according to the spectral type K2, the estimated mass is $ M_{2}=0.73M_{\bigodot_{}^{}}$, (\citealt{Cox+2000}). And the mass of primary component calculated as  $M_{1}=0.35 M_{\bigodot_{}^{}}$. Our solution confirmed that KIC 9026766 is a near-contact binary with a fill-out factor of $0.2$ percent.

In the second part, we studied the O-C curves of minima to investigate the orbital variation. This target is flagged as TM on the Kepler website, so it was feasible to investigate the presence of a third body. By fitting a quadratic function to the O-C curve of the primary minima, we calculated a period variation of $ \dot{P}=1.784911\times10^{-9} \pm7.08752\times10^{-12}\frac{day}{year}$. Moreover, we inferred the third body with a period of $ 972.5866\pm0.0041$ days and the mass of $ M=0.029M_{\bigodot_{}^{}}$.

The modulation period for the difference between the maxima and the main peak of periodogram analysis of residuals of the third-body effect as well as the observed frequency peak for the relative luminosity of the primary and secondary eclipses have the same values. The Applegate mechanism and spot motion cannot explain the existence of this signal with a period of just more than three times the orbital period. Background effects are possible because of the two nearby stars Gaia DR2 2127972818460821504 and Gaia DR2 212 7972818467257344. 
For the rest of the signals in the periodogram of the O-C curve of minima, the spot evolution was considered and the values of the amplitude and periods of the signals implied the effect of spot motion on the system components.

\begin{acknowledgements}
The authors would like to thank Kyle Conroy for helpful discussions. This paper has made use of data from the Kepler mission. Funding for the Kepler mission is provided by the NASA Science Mission directorate. This research has made use of SIMBAD and VIZIER databases, operated at CDS, Strasbourg, France. This research has made use of Lightkurve, a Python package for Kepler and TESS data analysis.
\end{acknowledgements}

\label{lastpage}

\end{document}